\documentclass[a4paper,11pt]{article}
\usepackage{jheppub} % for details on the use of the package, please see the JINST-author-manual
\usepackage{lineno}
\usepackage{xcolor}
\newcommand{\preprintnumber}{DESY-26-099}
\arxivnumber{2607.22390} % if you have one

\title{\boldmath Simultaneous extraction of top quark mass, strong coupling, effective mixing angle, and proton PDFs using inclusive DIS and proton-proton collision data.}

\author{Mikel~Mendizabal Morentin$^{1}$, Katerina Lipka$^{1,2,3}$, Giovanni Limatola$^{1,4}$ Efstathios Agrafiotis$^{1}$, Oleksandr Zenaiev$^4$, XiaoMin Shen$^{5,6}$, Sven Olaf Moch$^4$.}

\affiliation{$^1$Deutsches Elektronen-Synchrotron DESY, Notkestr. 85, 22607 Hamburg, Germany}

\affiliation{$^2$Fakult{\"a}t f{\"u}r Mathematik und Naturwissenschaften, Bergische Universit{\"a}t Wuppertal, Gau{\ss}strasse 20, D-42119 Wuppertal, Germany}

\affiliation{$^3$Fysiska institutionen, Lunds universitet, Lund; Sweden}

\affiliation{$^4$II. Institute for Theoretical Physics, University of Hamburg, 
Luruper Chaussee 149, 22761 Hamburg, Germany}
\affiliation{$^5$Institute of Modern Physics, Chinese Academy of Sciences, Lanzhou, Gansu 730000, China}
\affiliation{$^6$University of Chinese Academy of Sciences, Beijing 100049, China}

\emailAdd{mikel.mendizabal.morentin@desy.de}
\emailAdd{katerina.lipka@desy.de}
\emailAdd{xiaominshen@impcas.ac.cn}
\emailAdd{oleksandr.zenaiev@desy.de}
\emailAdd{giovanni.limatola@desy.de}
\emailAdd{sven-olaf.moch@desy.de}
\emailAdd{efstathios.agrafiotis@desy.de}

\abstract{We present the first simultaneous determination of the proton parton distribution functions (PDFs) together with the strong coupling \(\alpha_s(m_Z)\), the top quark pole mass \(m_t^{\rm pole}\), and the effective weak mixing angle \(\sin^2\theta_{\text{eff}}\). 
The analysis is performed at next-to-next-to-leading order in QCD using the \texttt{xFitter} framework, based on inclusive deep-inelastic scattering measurements at HERA and precise measurements of jet, electroweak boson, and top quark-antiquark pair production in proton-proton collisions at the LHC. Non-relativistic QCD effects in top quark-antiquark pair production are considered at next-to-leading order, and combined with next-to-leading logarithmic soft-gluon resummation. 
Simultaneously with PDFs and Standard Model (SM) parameters, relevant top quark SM effective-field-theory (SMEFT) operators are constrained. The precision of the obtained SM parameters is competitive with the state-of-the art individual extractions and accounts for the correlations among those parameters and PDFs. This work paves the way for future global interpretation of the LHC data in terms of QCD, SM and SMEFT. }

\begin{document}
\begin{flushright}
\preprintnumber
\end{flushright}
\maketitle

\flushbottom

\section{Introduction}
\label{sec:intro}
High-energy hadronic collisions are described through the QCD factorisation theorem, which separates the long-distance, non-perturbative structure of the hadron from the short-distance, hard-scattering process at a factorisation scale $\mu_f$.
While the hard-scattering cross-section is calculated using perturbative QCD (pQCD), the non-perturbative proton structure is encoded in parton distribution functions (PDFs) $f(x)$ and is determined from global QCD analyses formulated as an inverse problem of factorisation. Therefore not only pQCD predictions, but also the PDFs depend on the values of the fundamental parameters of the Standard Model (SM) at a certain scale, such as particle masses and coupling strengths. The convolution of these ingredients in the QCD factorisation formula creates an intrinsic correlation between the extracted PDFs and the SM parameters, making a simultaneous determination essential for a consistent and unbiased interpretation of LHC data. 
Furthermore, such correlations may weaken constraints on higher-dimensional operators in the Standard Model Effective Field Theory (SMEFT), which are inferred from deviations of measured spectra from SM predictions in hadronic collisions. 
In this work, we present a simultaneous extraction of PDFs, strong coupling at the $Z$ boson mass $\alpha_S(m_Z)$, the mass of the top quark $m_t$ and the effective electroweak mixing angle $\sin^2\theta_{\text{eff}}^{f}$, considering the correlations of these parameters.  

The electroweak mixing angle \(\theta_W\) is a fundamental parameter of the SM that describes the relation between the masses of the electroweak bosons
\begin{equation}
\sin^{2}{\theta_{W}} =  1-\frac{m_{W}^2}{m_{Z}^2},
\end{equation}
% csm NO !!
%
where \(m_Z\) and \(m_W\) are the masses of the $Z$ and $W$ bosons, respectively. At leading order (LO) in the electroweak (EW) interaction, it also determines the vector (\(v_f\)) and axial (\(a_f\)) couplings of the $Z$ boson to  the fermion $f$:
\begin{equation}
\frac{v_{f}}{a_{f}} = 1 - 4|Q_{f}|\sin^{2}{\theta_{W}},
\end{equation}
where $Q_f$ is the electric charge of the fermion $f$, in units of the electric charge of the electron $e$.
At higher orders, EW radiative corrections modify this simple relation. This necessitates the introduction of an effective, flavour-dependent electroweak mixing angle,
\begin{equation}
\sin^2\theta_{\text{eff}}^{f} = \kappa_{f}\sin^{2}{\theta_{W}},
\end{equation}
with \(\kappa_f\) representing the process- and flavour-dependent corrections in a given renormalization scheme.
Although the most precise single measurements of \(\sin^2\theta_{\text{eff}}\) were performed by LEP and SLD at the $Z$ resonance~\cite{ALEPH:2005ab}, the recent result~\cite{CMS:2024ony} obtained by the CMS experiment at the LHC, \(\sin^2\theta_{\text{eff}}^{\ell} = 0.23152 \pm 0.00031\)~\cite{CMS:2024ony}, brings the precision in $pp$ collisions to similar level. Still,  the measurements using data collected in $pp$ collisions are limited by the proton PDFs in two ways: the choice of a central PDF set leads to a shift in the central value of $\sin^2\theta_{\text{eff}}$ and the PDF uncertainty is the dominant source of uncertainty in the final result. 

%\textcolor{red}{KL: Why do we need it? EW fit assumes only SM, while we do individual measurement here... The most precise extraction of \(\sin^2\theta_{\text{eff}}\) comes from a global EW fit~\cite{EWFit}, yielding \(0.23155 \pm 0.00004\).}

The mass of the top quark $m_t$ has a special role in the Standard Model because of its importance in both electroweak and strong sectors of the SM Lagrangian. %Different approaches to extract $m_t$ lead to different level of precision in its value and the accuracy of its interpretation, as detailed in the exhaustive review~\cite{CMS:2024irj}. 
Different approaches to extracting $m_t$ yield varying levels of precision and interpretive accuracy, as detailed in the exhaustive review~\cite{CMS:2024irj} and references therein.
While the so-called direct measurements from the kinematic fit have high precision \(172 \pm 0.27\) GeV~\cite{ParticleDataGroup:2026aaa}, the uncertainty in the mass of the top quark, obtained in a well-defined renormalisation scheme, e.g. pole mass, $m_t^{\rm pole}$, is still of the order of about 1 GeV. Furthermore, the $m_t^{\rm pole}$ extractions are mostly based on the invariant mass $M_{t\bar{t}}$ of the $t\bar t$ system close to the production threshold $M_{t\bar{t}}\simeq 2m_t$, where NRQCD effects play a relevant role. Consideration of the latter in description of $t\bar t$ production at the LHC is further motivated by the recent observations~\cite{CMS:2025kzt,ATLAS:2026dbe} of an excess at $t\bar t$ production threshold, which hint to essential sensitivity of the LHC measurements to NRQCD effects. 
A recent work by the NNPDF group~\cite{Ball:2026qno} accounts for such NRQCD effects, based on the simulation~\cite{Fuks:2024yjj}, and reports a resulting shift in the extracted $m_t$ by +0.6~GeV. In our work,  the incorporated NRQCD effects are calculated at next-to-leading order (NLO) in Ref.~\cite{Garzelli:2024uhe}, and also consider next-to-leading logarithmic (NLL) soft-gluon threshold resummation.

The strong coupling constant $\alpha_s$ is the least well known SM coupling, governing both the evolution of PDFs and the dynamics of partonic interactions. Several simultaneous determinations of PDFs and $\alpha_s(m_Z)$ exist, provided by global QCD groups as well as by the LHC experimental works.
For instance, the CTEQ-TEA group reports the value of $\alpha_s(m_Z)=0.1164\pm0.0026$~\cite{Hou:2019efy}, while the result by the ABMP group is  $\alpha_s(m_Z)=0.1152\pm 0.0008$~\cite{Alekhin:2025qdj}
%csm 
%\textcolor{blue}{SZ One might want to cite the latest work \url{https://arxiv.org/abs/2510.21435} which report $0.1152 \pm 0.0008$. Sven, what do you think?} 
at next-to-next-to-leading order (NNLO) in QCD. 
In such analyses, the $\alpha_s$-sensitivity of jet and $t\bar t$ production in $pp$ collisions is explored, also investigated in experimental analyses, e.g. Refs.~\cite{CMS:2019esx,CMS:2021yzl,CMS:2024trs}. 

In this work, we perform a simultaneous extraction of PDFs together with these SM parameters, which allows us to study and mitigate the correlations between them. The present analysis is based on minimal selection of deep-inelastic scattering (DIS) and proton-proton ($pp$) collision data, summarized in Table~\ref{tab:datasets}, providing sensitivity to the parameters of interest. 

\begin{table}[htbp]
\centering
\begin{tabular}{l|c|c|c}
\textbf{Data set} & \textbf{Process / Observable} & \textbf{$\sqrt{s}$} & \textbf{Reference} \\ \hline
$ep$ DIS & $ep \to eX$: NC+CC inclusive & 0.318 TeV & arXiv:1506.06042~\cite{H1:2015ubc} \\
 & cross sections (HERA I+II) & & \\ \hline
$pp$ jets & $pp \to \text{jet}+X$ & 2.76 TeV & arXiv:1512.06212~\cite{CMS:2015jdl} \\
 & $d\sigma/dp^{\rm jet}_{T}d|Y^{\rm jet}|$ (R=0.7) & 7 TeV & arXiv:1410.6765~\cite{CMS:2014qtp} \\
 & & 8 TeV & arXiv:1609.05331~\cite{CMS:2016lna} \\
 & & 13 TeV & arXiv:2111.10431~\cite{CMS:2021yzl} \\ \hline
$pp$ DY $A_4$  & $pp \to Z/\gamma^* \to \ell^+\ell^-$: $A_{4}$ & 13 TeV & arXiv:2408.07622~\cite{CMS:2024ony} \\
 &  & & \\ 
$pp$ $W^\pm$ asymmetry & $pp \to W^\pm \to \ell^\pm \nu$: & 13 TeV & arXiv:2008.04174~\cite{CMS:2020cph} \\
 & $d\sigma/d|\eta_{\ell}|$ & & \\ \hline
$pp$ $t\bar t$ ($\ell$+jet) & $pp \to t\bar{t}$ ($\ell$+jet) & 13 TeV & arXiv:2108.02803~\cite{CMS:2021vhb} \\
 & $d\sigma/dm_{t\bar{t}}$  & & \\
$pp$ $t\bar t$ ($ 2\ell$) & $pp \to t\bar{t}$ ($2\ell$) & 13 TeV & arXiv:2402.08486~\cite{CMS:2024ybg} \\
 & $d\sigma/dm_{t\bar{t}}$ & & \\\hline
\end{tabular}
\caption{Summary of the experimental measurements and their associated theoretical predictions used in the global analysis.}
\label{tab:datasets}
\end{table}

The DIS measurements at HERA form the basis of the PDF determination, constraining the gluon distribution indirectly through scaling violations. In contrast, jet and top-antitop pair production in $pp$ collisions at the LHC directly probe the gluon distribution and provide sensitivity to  $m_t^{\rm pole}$ and $\alpha_s$, while electroweak boson production constrains the light-quark distributions and $\sin^2\theta_{\text{eff}}$.
As an example of extending the analysis to a global simultaneous SMEFT interpretation, we also investigate two higher-dimensional operators in $t\bar t$ production, following the approach of Ref.~\cite{Shen:2024sci}. In particular, we study the sensitivity of the $t\bar t$ measurements at the LHC to the Wilson coefficients $c_{tG}$ and $c_{tq}^{(8)}$, which in standard SMEFT interpretations are affected by correlations with the proton PDFs and the value of  $m_t^{\rm pole}$.

The paper is structured as follows: in Section~\ref{sec:measurements}, the experimental measurements used in the interpretation are introduced and their corresponding theoretical predictions are described. Section~\ref{sec:nrqcd} is dedicated to the calculation of NRQCD effects in $t\bar t$ production in $pp$ collisions. In Section~\ref{sec:analysis}, the analysis strategy and the fit framework are presented. In Section~\ref{sec:results}, the extracted SM parameters, PDFs, and the SMEFT Wilson coefficients are discussed. The work is summarized in Section~\ref{sec:conclusions}.

\section{Data sets and respective theory predictions}
\label{sec:measurements}
The experimental measurements 
utilised in this work are selected according to sensitivity to an individual or several parameters of interest. 
While the most comprehensive analysis of all available relevant LHC measurements is a subject of future studies, a minimal set of high-sensitivity and highest-precision data is chosen for this work as a proof of principle. All published systematic correlations among and across the HERA and the LHC data are considered. 

The baseline dataset for the PDF determination is the combination of inclusive neutral-current (NC) and charged-current (CC) deep-inelastic scattering (DIS) measurements from the H1 and ZEUS experiments at HERA~\cite{H1:2015ubc}. Theoretical predictions for NC and CC DIS are generated within the \texttt{xFitter} framework using \texttt{APFEL} QCD evolution in the FONLL general-mass variable-flavour-number scheme (GM-VFNS) at NNLO accuracy~\cite{Forte:2010ta},
see also \cite{Ablinger:2025joi,Ablinger:2025awb} for recent work on the VFNS at three loops in pQCD. Since the HERA data are sensitive to low-\(x\) resummation and higher-twist effects at low \(Q^2\), a kinematic cut of \(Q^{2} > 10\) GeV\(^2\) is applied to ensure the precision of fixed-order pQCD calculations~\cite{Bertone:2022sso,Bertone:2024snr,xFitterResummationScale}.

Inclusive jet production in $pp$ collisions is particularly sensitive to both $\alpha_s$ and PDFs. We use the CMS double-differential measurements of inclusive jet production cross-section as a function of jet transverse momentum \(p_T\) and absolute rapidity \(|y|\)
at centre-of-mass energies of \(\sqrt{s}=2.76\)~\cite{CMS:2015jdl}, 7~\cite{CMS:2012ftr,CMS:2014qtp}, 8~\cite{CMS:2016lna}, and 13~\cite{CMS:2021yzl} TeV, with recently reevaluated systematic correlations, provided in Ref.~\cite{CMS:2024trs}. 
Inclusive jet production is a powerful probe of \(\alpha_s\) already at LO, while the different \(\sqrt{s}\) provide access to complementary regions of $x$ probing different PDFs at NNLO, as demonstrated in Ref.~\cite{CMS:2024trs}. The theoretical predictions are computed with \texttt{NNLOJET}~\cite{Currie:2016bfm} at NNLO accuracy in the leading-colour approximation, and are interfaced to \texttt{APPLfast}~\cite{Britzger:2022lbf} interpolation grids. The renormalisation $\mu_r$ and factorisation $\mu_f$ scales are set to the transverse momentum of the individual jet. The predictions are corrected for non-perturbative and NLO EW effects.

Several Drell-Yan (DY) measurements in $pp$ collisions at \(\sqrt{s}=13\) TeV are included, in both NC and CC channels. For NC DY, we incorporate the CMS measurement~\cite{CMS:2024ony} of the $A_4$ angular coefficient as a function of the dilepton invariant mass and absolute rapidity. This observable provides direct sensitivity to \(\sin^2\theta_{\text{eff}}\). Additionally, NC DY data provides important constraints on the valence quark distributions. The $A_4$ predictions are computed at NLO in QCD using \texttt{MC@NLO}, employing the \(G_F\) EW scheme, and are supplemented by NNLO QCD and NLO EW correction factors. For CC DY, we include the CMS measurement of the W boson lepton rapidity asymmetry~\cite{CMS:2020cph}, which offers sensitivity to the sea quark distributions. Corresponding theoretical predictions are computed using \texttt{MATRIX}~~\cite{Grazzini:2017mhc,Catani:2019hip} interfaced with \texttt{PineAPPL}~\cite{Carrazza:2020gss, Jezo:2026adf} grids, at NNLO in QCD and NLO in EW accuracy, using the \(G_F\) scheme and a complex-mass scheme for the EW bosons.

Finally, we incorporate two CMS measurements of \(t\bar{t}\) production as a function of the invariant mass $M_{t\bar t}$ of the $t\bar{t}$ system at $\sqrt{s}=13$ TeV in lepton+jet~\cite{CMS:2021vhb} and in the dilepton~\cite{CMS:2024ybg} final states. These measurements are particularly sensitive to the value of $m_t^{\rm pole}$. Similar measurement by the ATLAS Collaboration~\cite{ATLAS:2019hxz} was investigated but finally not considered in the final fit due to its lower sensitivity to parameters of interest. Inclusion of this measurement does not change any conclusion of this work.

The theoretical predictions are produced with \texttt{MATRIX}~\cite{Grazzini:2017mhc, Catani:2019iny, Catani:2019hip} interfaced~\cite{Garzelli:2023rvx} with \texttt{PineAPPL} at NNLO accuracy in QCD \textcolor{blue}, with a dynamical scale choice
\begin{equation}
\label{eq:mtt_scale}
    \mu_{R}=\mu_{F} = \frac{1}{4}\left(\sqrt{m^2_{t,1}+p^2_{T,1}} + \sqrt{m^2_{t,2}+p^2_{T,2}}\right),
\end{equation}
where $m_{t,1 (2)}, p_{T,1(2)} $ are the mass and the transverse momentum of the top quark (1) and of the top antiquark (2), respectively.
To accurately model the $M_{t\bar t}$ threshold region, we include the effects of NRQCD and soft-gluon resummation~\cite{Kiyo:2008bv,Garzelli:2024uhe,Garzelli:2026ctb} as an effective correction to the fixed-order (FO) predictions. The calculation of these NRQCD corrections is detailed in the following section.

%A similar measurement by the ATLAS Collaboration in the lepton+jets final state was also considered. However, its sensitivity to small variations in \(m_t\) in the threshold region was found to be negligible compared to the CMS measurements. 
%We quantify this in Figure~\ref{fig:mt_sensitivity}, which displays the sensitivity for each measurement. The sensitivity in bin \(i\) is defined as
%\begin{equation}
%    S_{i} = \frac{|T_{\Delta,i}-T_\mathrm{0,i}|}{\sigma_{i}},
%\end{equation}
%where \(T_{0,i}\) is the central theory prediction for \(m_t=172.5\) GeV, \(T_{\Delta,i}\) is the prediction with a shift of \(\Delta m_t = \pm 0.5\) GeV, and \(\sigma_i\) is the measurement uncertainty. The CMS measurements are found to be roughly twice as sensitive as the ATLAS measurement to a given change in \(m_t\), particularly in the threshold region.
%\begin{figure}[htbp]
%    \centering    \includegraphics[width=0.5\linewidth]{figures/mt_sensitivity.pdf}
 %   \caption{Sensitivity of the CMS and ATLAS \(m_{t\bar{t}}\) measurements to a variation of \(\Delta m_t = \pm 0.5\) GeV in the top quark mass. The CMS measurements demonstrate approximately double the sensitivity of the ATLAS measurement, justifying their inclusion in the analysis.}
%\label{fig:mt_sensitivity%}
%\end{figure}

\section{Non-relativistic QCD corrections to $t\bar t$ production in $pp$ collisions}
\label{sec:nrqcd}

\begin{figure}[htbp]
    \centering
    \includegraphics[width=0.7\linewidth]{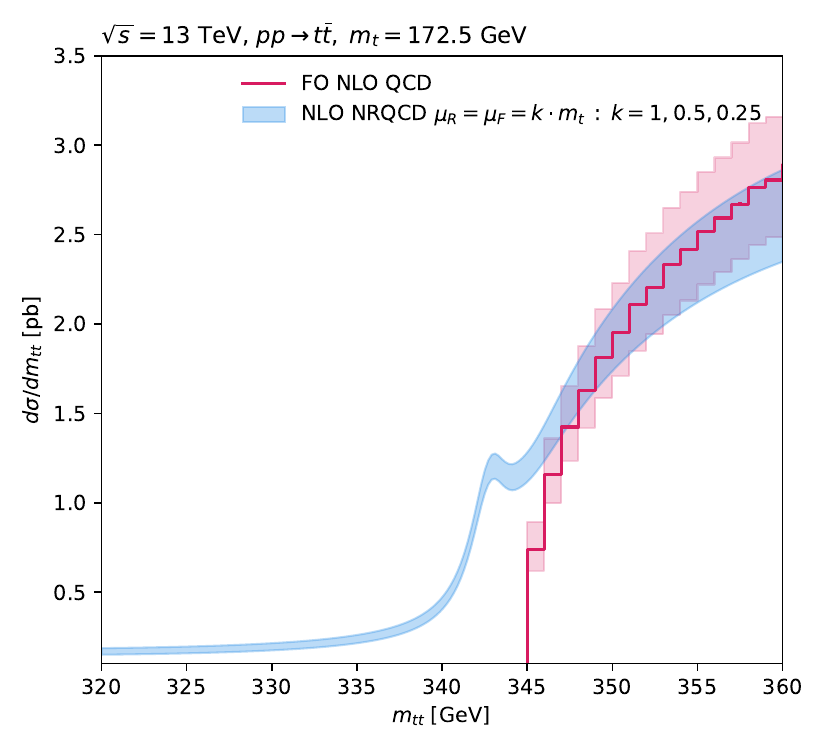}
    \caption{Comparison of the fixed-order NLO QCD prediction (red) and the NLO NRQCD prediction (blue) for the differential cross-section as a function of \(M_{t\bar{t}}\). The bands represent the scale uncertainties. The validity region of the NRQCD calculation is defined by the intersection of the central values and its scale-uncertainty bands.}
    \label{fig:mtt_nrqcd_nloqcd}
\end{figure}

Since in the threshold region $M_{t\bar{t}}\simeq 2m_t$ the produced top quarks move with  velocity $\beta_t=\sqrt{1-4m_t^2/M_{t\bar{t}}^2}\sim\alpha_s\ll 1$, in the $t\bar{t}$ centre-of-mass frame, the cross section experiences significant QCD corrections due to low-energy quasi-bound-state interactions, dominated by Coulomb-like corrections  scaling as $(\alpha_s/\beta_t)^n$ at $n$ loops, which need all-order resummation. 
This can be achieved working within the framework of effective field theory Non-Relativistic QCD (NRQCD)~\cite{Lepage:1992tx,Pineda:1997bj}, which allows the factorisation of $(\alpha_s/\beta_t)^n$ corrections into Green's functions which solve Schrödinger equations for QCD Coulomb-like potentials $V_{\rm{QCD}}$.
The presence of such quasi-bound state effects in heavy-quark pair production is well established within the SM, and has been extensively studied in the past for the $t\bar{t}$ case, both in lepton collision~\cite{Bigi:1986jk,Fadin:1987wz}, where the $t\bar{t}$ pair is produced in a colour-singlet final state, and in hadronic collisions~\cite{Fadin:1990wx,Hagiwara:2008df,Kiyo:2008bv,Ju:2020otc}, where  a mixture of colour-singlet and octet final state must be considered. 
In this work we will consider  NRQCD effects involving $t\bar{t}$ pair production at the LHC in the threshold region, building on the results in~\cite{Kiyo:2008bv,Garzelli:2024uhe,Garzelli:2026ctb}, where differential predictions in $M_{t\bar{t}}$ have been provided at NLO in NRQCD, also accounting for the resummation of logarithmic-enhanced  terms associated with soft-gluon emissions, up to NLL accuracy. In the absence of a formal matching prescription to combine NRQCD effects, essential in the threshold region, with a usual fixed-order (FO) pQCD computation which describes the large $M_{t\bar{t}}$ region, also called \emph{continuum}, in this study  
 we incorporate the NRQCD effects as an additive correction to the FO calculation. Relying on the updated cross-section prescriptions of~\cite{Garzelli:2024uhe, Garzelli:2026ctb} we estimated the validity region of the NRQCD approximation in  $|M_{t\bar{t}}-2m_t|\le 5~\rm{GeV}$, identifying the region where the NRQCD predictions and the NLO pQCD results start to overlap. The central renormalisation and factorisation scales are set to $\mu_r=\mu_f=m_t/2$, which is the value approached by Eq.~\ref{eq:mtt_scale} in the threshold region, where $p_{T,t(\bar{t})} \rightarrow 0$. In  Figure~\ref{fig:mtt_nrqcd_nloqcd} we show the $M_{t\bar{t}}$ differential distribution at NLO in NRQCD, together with the FO NLO predictions, from which it is evident that the two curves  begin to overlap for \(M_{t\bar{t}} \approx 355\) GeV.
We therefore define the cross section excess $\Delta\sigma_{\rm{NRQCD}}$, arising from quasi-bound-state effects in the threshold region, as the difference between the  NLO NRQCD $M_{t\bar{t}}$ distribution ($\sigma_{\rm{NRQCD}}$), and the FO NLO prediction ($\sigma_{\rm{NLO}}$), integrated over the $M_{t\bar{t}}\in[340,355]~\rm{GeV}$ bin, obtaining
\begin{equation}
\label{eq:deltanrqcd}
    \Delta\sigma_\mathrm{NRQCD}(340~{\rm{GeV}}\le M_{t\bar{t}}\le 355~{\rm{GeV}})= \sigma_\mathrm{NRQCD} - \sigma_\mathrm{NLO} =  6.04^{+1.93}_{-0.91}~\mathrm{\;pb},
\end{equation}
with the uncertainties estimated by varying simultaneously $\mu_r=\mu_f$ by a factor of two in the range $[m_t/4,m_t]$. Since the estimated region of validity of the NRQCD contribution affects only the first bin of the \(M_{t\bar{t}}\) distribution for both CMS measurements we will treat \(\Delta\sigma_\mathrm{NRQCD}\) as an additive factor to the theoretical prediction in this first bin. 
Furthermore, we evaluate $\Delta\sigma_{\rm{NRQCD}}$ also including NLL soft-gluon threshold resummation effects on the NRQCD distribution, according to~\cite{Kiyo:2008bv, Garzelli:2024uhe}. This is found to further enhance the cross section excess in Eq.~\ref{eq:deltanrqcd} by about 10\%, resulting in
\begin{equation}
    \Delta\sigma_\mathrm{NRQCD+NLL}= \sigma_\mathrm{NRQCD+NLL} - \sigma_\mathrm{FO} =  7.31^{+1.93}_{-0.91  } \mathrm{\;pb}.
\end{equation}

\begin{figure}[htbp]
    \centering
    \includegraphics[width=0.7\linewidth]{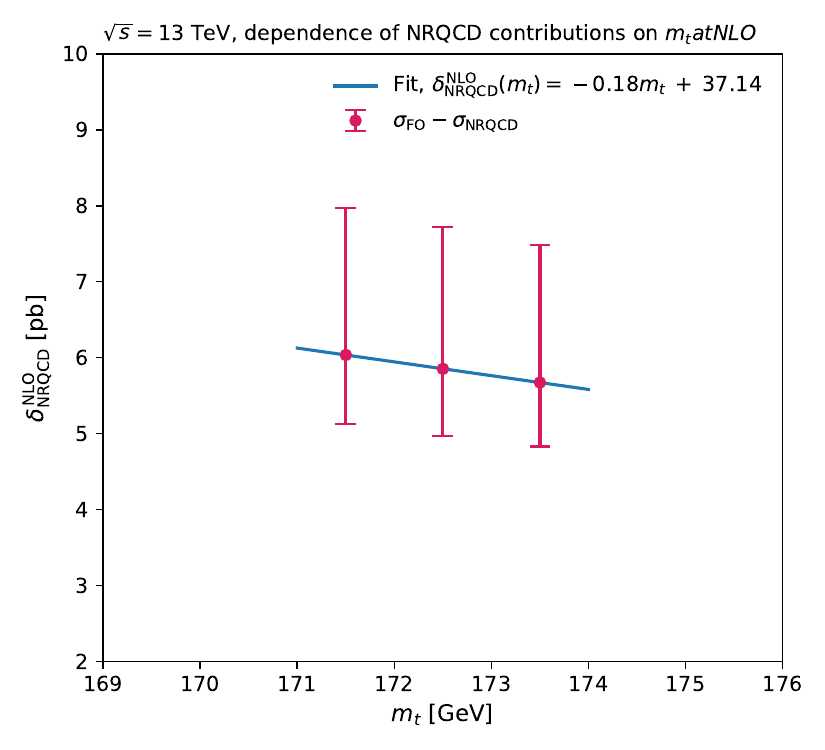}
    \caption{The additive NRQCD correction, \(\Delta\sigma_\mathrm{NRQCD}\), as a function of the top quark mass. The observed mass dependence is negligible compared to the scale uncertainty of the NRQCD calculation.}
    \label{fig:deltanrqcd_mt}
\end{figure}

The dependence of this additive factor on \(m_t\) is also examined and shown in Figure~\ref{fig:deltanrqcd_mt}. While a slight mass dependence is observed, it is found to be well within the scale uncertainty of the NRQCD calculation. Consequently, we treat \(\Delta\sigma_\mathrm{NRQCD}\) as independent of \(m_t\).
% We need to be very clear on the way we add the NRQCD correction. We can only extract the NRQCD excess w.r.t. NLO pQCD results, and then add this K-factor on top of the NNLO distribution.
%The point is only to clarify that the upper bound of the integration range is estimated by looking at the region where the two NLO curves begin to overlap.  In this sense we are assuming that our K-factor is only valid in the [340,355] GeV range, and that the upper bound stays approximately the same when going up to NNLO.

\section{The QCD+EFT analysis}
\label{sec:analysis}

The QCD (QCD+SMEFT) fit is performed using the \texttt{xFitter} framework~\cite{Alekhin:2014irh,xFitterDevelopersTeam:2017xal}, which allows for the simultaneous extraction of $\alpha_s(m_Z)$, $m_t^{\rm pole}$, and $\sin^2\theta_{\text{eff}}$ together with the PDFs at NNLO in QCD. The minimisation of the $\chi^2$ function is carried out using the \texttt{Ceres-Solver}~\cite{Agarwal_Ceres_Solver_2022}.

\subsection{PDF parametrisation}

The final PDF parametrisation is obtained as a result of the parametrisation scan, described in the following. At this stage, the SM parameters \(\alpha_s(m_Z)\) and \(m_t\) are kept fixed. The following assumptions are made. The starting scale for the DGLAP evolution is set to $Q_0^2 = 10$ GeV$^2$, consistent with the kinematic cut on the DIS data. The charm and bottom quark masses are fixed to $m_c = 1.47$ GeV and $m_b = 4.5$ GeV, respectively. The charm threshold factor, $k_{m_c}=2.16$, is chosen such that the charm quark threshold coincides with the starting scale, thereby effectively neglecting an intrinsic charm component.

At the starting scale, the gluon, \(xg(x)\), the up and down valence distributions, \(xu_v(x)\) and \(xd_v(x)\), the anti-up, anti-down, and strange sea distributions, \(x\bar{u}(x)\), \(x\bar{d}(x)\), and \(xs(x)\) are parametrized. The functional form is based on the Bonvini-Giuli parametrisation~\cite{Bonvini:2019wxf}:
\begin{equation}
    xf(x) = A_{f}x^{B_{f}}\left(1-x\right)^{C_{f}}\left(1+D_{f}x+E_{f}x^{2}+F_{f}\log x+G_{f}\log^{2}x\right).
\end{equation}
To determine the optimal parametrisation form, we begin by setting the \(D_f, E_f, F_f\), and \(G_f\) parameters to zero. Additional parameters are then included in the fit one by one if they yield a significant improvement in the \(\chi^2\). The final parametrisation at the starting scale is:
\begin{align}
    xg(x) &= A_{g}x^{B_{g}}(1-x)^{C_{g}}(1+E_{g}x^{2}+ F_{g}\log x),\\
    xu_{v}(x) &= A_{u_{v}}x^{B_{u_{v}}}(1-x)^{C_{u_{v}}}(1+D_{u_{v}}x^{2}),\\
    xd_{v}(x) &= A_{d_{v}}x^{B_{d_{v}}}(1-x)^{C_{d_{v}}},\\
    x\bar{u}(x) &= A_{\bar{d}}x^{B_{\bar{d}}}(1-x)^{C_{\bar{u}}}(1+F_{\bar{u}}\log x),\\
    x\bar{d}(x) &= A_{\bar{d}}x^{B_{\bar{d}}}(1-x)^{C_{\bar{d}}}(1+D_{\bar{d}}x+F_{\bar{d}}\log x),\\
    xs(x) &= A_{\bar{d}}\frac{f_{s}}{1-f_{s}}x^{B_{\bar{d}}}(1-x)^{C_{\bar{d}}},
\end{align}
where \(f_s\) is the strangeness fraction, which is fixed to \(f_s = 0.4\) following the HERAPDF2.0 analysis~\cite{H1:2015ubc}.

\subsection{Extraction of the SM parameters}

Based on the optimal PDF parametrisation found, the SM parameters are freed in the fit, which leads to reevaluation of the PDF parameters as well.  While \(\alpha_s(m_Z)\) can be varied directly in the interpolation grids and DGLAP evolution, the values of $m_t^{\rm pole}$ and \(\sin^2\theta_{\text{eff}}\) require technically different approach. For $m_t^{\rm pole}$, the \texttt{EFTREACTION} module~\cite{Shen:2024sci} of \texttt{xFitter} is used, which linearly interpolates the cross-section as function of $m_t$ in the region of interest,
\begin{equation}
    \sigma(m_t + \delta m_t) = \sigma(m_t) + \sigma_{\delta m_t} \times \delta m_t,
\end{equation}
where \(\sigma_{\delta m_t}\) is the linear coefficient.
%, and \(\delta m_t = m_t - 172.5\) GeV. 
This interpolation has been shown to be sufficient for the kinematic range under consideration~\cite{Shen:2024sci}. For \(\sin^2\theta_{\text{eff}}\), we adopt the approach of~\cite{CMS:2024ony}, introducing it as a theory nuisance parameter in the \(\chi^2\) calculation, profiled only for the DY $A_4$ measurement, as the other datasets (HERA, jets, W asymmetry) show negligible sensitivity to it.

\subsection{SMEFT parameters}

Furthermore, the analysis is extended to explore the effects of new physics by including selected dimension-6 SMEFT operators. We focus on two operators relevant for \(t\bar{t}\) production, \(O_{tG}\) and \(O_{tq}^{(8)}\), defined in the Warsaw basis~\cite{Grzadkowski:2010es} with corresponding Wilson coefficients \(c_{tG}\) and \(c_{tq}^{(8)}\). The cross-section dependence is assumed to be quadratic, following the approach of~\cite{Shen:2024sci}:
\begin{equation}
    \sigma\left(c_{tG},c_{tq}^{(8)}\right) = \sigma_{SM} + \sum_{i=\left\{c_{tG},c_{tq}^{(8)}\right\}} \frac{c_i}{\Lambda^2}\sigma_i + \sum_{i,j=\left\{c_{tG},c_{tq}^{(8)}\right\}} \frac{c_ic_j}{\Lambda^4}\sigma_{i,j},
\end{equation}
where \(\sigma_i\) and \(\sigma_{ij}\) are the linear and quadratic contributions to the cross-section, respectively, and \(\Lambda\) is the SMEFT scale. The SMEFT contributions are implemented via the \texttt{EFTREACTION} module using \texttt{PineAPPL} interpolation grids. These grids were produced at NLO in QCD using \texttt{MadGraph5\_aMC@NLO}~\cite{Alwall:2014hca}.
% with benchmark values of \(c_{tG}/\Lambda^2=20 \text{ TeV}^{-2}\) and \(c_{tq}^{(8)}/\Lambda^2=80 \text{ TeV}^{-2}\). 
Two grids are generated for each coefficient to account for the linear and quadratic contributions. The mixing term between the two coefficients (\(c_{tG}\cdot c_{tq}^{(8)}\)) is neglected in this study.

\subsection{Uncertainties}
The sources of uncertainty are classified into three categories: experimental, model, and missing higher-order uncertainty (MHOU).

Experimental uncertainties account for the statistical and systematic uncertainties in the measurements, and are determined using \texttt{MINUIT}~\cite{James:1975dr} with the Pumplin method~\cite{Pumplin:2000vx,Pumplin:2002vw} using a tolerance \(\Delta\chi^2 = 1.0\), corresponding to a 68\% confidence level.

Model uncertainties arise from the choices made in the QCD analysis and are evaluated by varying the corresponding parameters. These variations include the minimum cut on DIS data (\(7.5 < Q^2_{\text{min}} < 12.5\) GeV\(^2\)), the PDF evolution starting scale (\(7.5 < Q_0^2 < 12.5\) GeV\(^2\)), the heavy quark masses (\(1.4 < m_c < 1.54\) GeV and \(4.25 < m_b < 4.75\) GeV), as well as the strangeness fraction (\(0.3 < f_s < 0.5\)).

The MHOU for the FO predictions are estimated by performing a 7-point variation of $\mu_r$ and $\mu_f$. While the factorisation scale $\mu_f$ is varied simultaneously for all the processes, the assumption on $\mu_r$ is varied independently for each process (jets, DY, and $t\bar t$ production). The final FO MHOU envelope is taken as the maximum and minimum deviations from the central fit results.  For the NRQCD additive factor, the MHOU is evaluated by varying its value in Eq.~\ref{eq:deltanrqcd} up and down. 

The total uncertainty is then computed by summing the experimental, model, FO MHOU, and NRQCD MHOU contributions in quadrature.

The extraction of uncertainty in $\sin^2\theta_{\mathrm{eff}}$ requires special treatment, since technically it is treated as a profiled nuisance parameter. To obtain the total experimental uncertainty, the 37 PDF replicas obtained from the fit are utilised. For each replica, the shift in $\sin^2\theta_{\mathrm{eff}}$ is profiled and then subtracted from the central shift. The resulting 36 shifts are summed in quadrature. Experimental uncertainties in $\alpha_s(m_Z)$ and $m_t$ are also propagated into the $\sin^2\theta_{\mathrm{eff}}$ determination. This is done by performing separate fits with $m_t$ (or $\alpha_s(m_Z)$) fixed to its value, varied by its uncertainty, and re-profiling $\sin^2\theta_{\mathrm{eff}}$. The resulting shifts on $\sin^2\theta_{\mathrm{eff}}$ are then added in quadrature.

\section{Results}
\label{sec:results}

In general, a very good agreement between the theory predictions and the
data is observed. The overall consistency across the experimental data is reflected in the quality of the fit, expressed in terms of 
partial $\chi^2$ divided by number of degrees of freedom, listed in Table~\ref{tab:dataset_chi2}, with obtained
global $\chi^2$ per degree of freedom of 1.05 including the correlated contribution of $41$.
\begin{table}[h!]
    \centering
    \begin{tabular}{|l|c|}
    \hline
        Measurement & $\chi^{2}/N_{\rm dp}$ \\\hline
        $ep$ DIS HERA I+II & 1244/1016  \\ \hline
        $pp$ inclusive jets 13~TeV & 79/78  \\
        $pp$ inclusive jets 8~TeV &  140/165 \\
        $pp$ inclusive jets 7~TeV & 74/130  \\
        $pp$ inclusive jets 2.76~TeV & 62/80  \\ \hline
        $pp$ DY $A_4$ 13~TeV & 62/63  \\
        $pp$ W asymmetry 13~TeV & 21/36  \\ \hline
        $pp$ $t\bar{t}$ l+ jets & 14/15  \\ 
        $pp$ $t\bar{t}$ 2l & 4/7  \\\hline 
        Total $\chi^2/N_{\rm dof}$ & 1658/1572 \\ \hline
    \end{tabular}
    \caption{The values of $\chi^2$ per number of data points $N_{\rm dp}$ for each individual data set as obtained in the fit, including contribution from correlated uncertainty sources, and the total $\chi^2$ per degree of freedom $N_{\rm dof}$. } 
\label{tab:dataset_chi2}
\end{table}

\begin{figure}[htbp]
    \centering
    \includegraphics[width=0.44\linewidth]{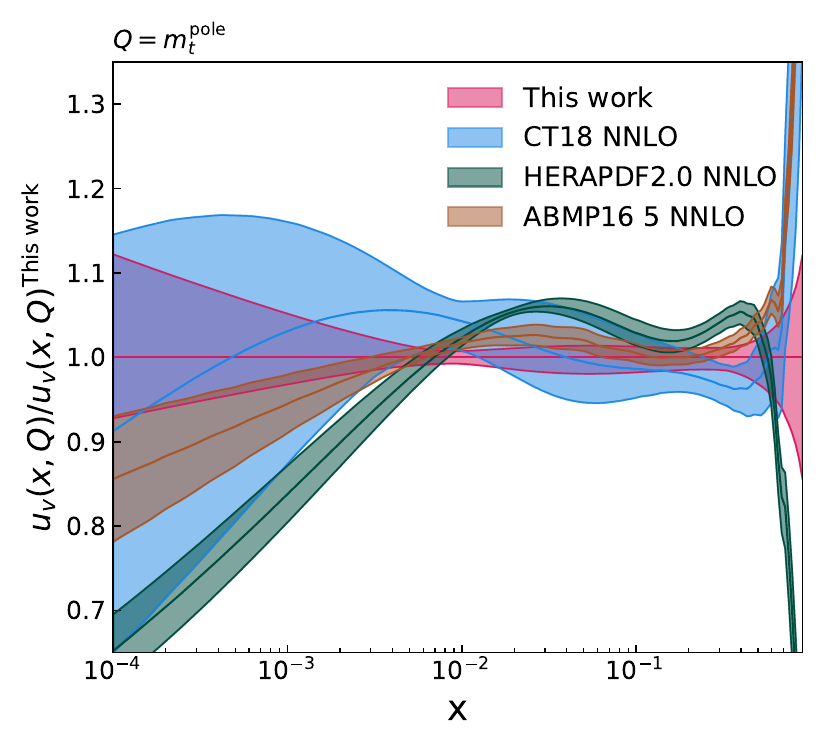}
    \includegraphics[width=0.44\linewidth]{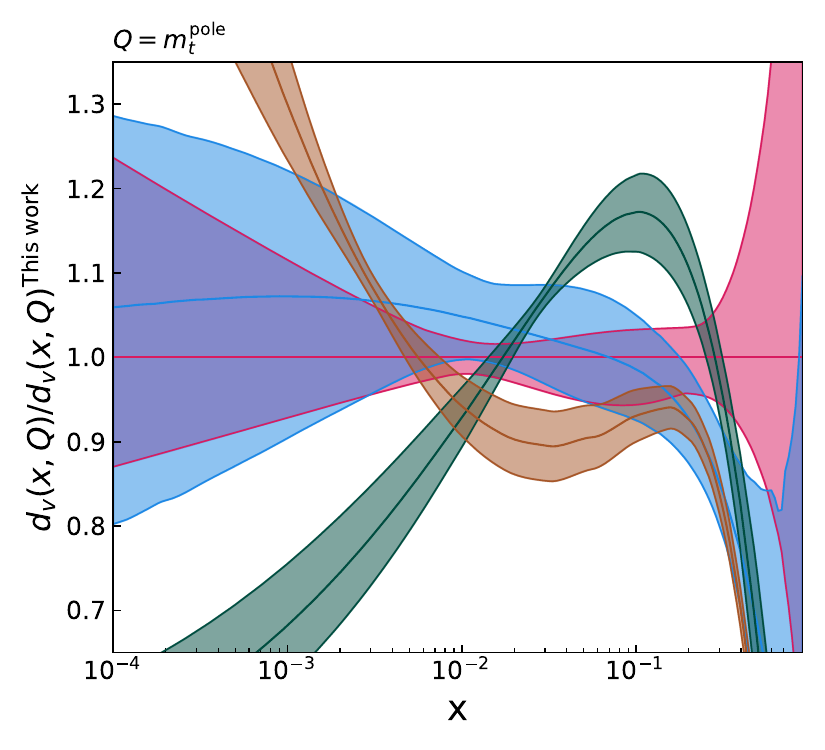}
    \includegraphics[width=0.44\linewidth]{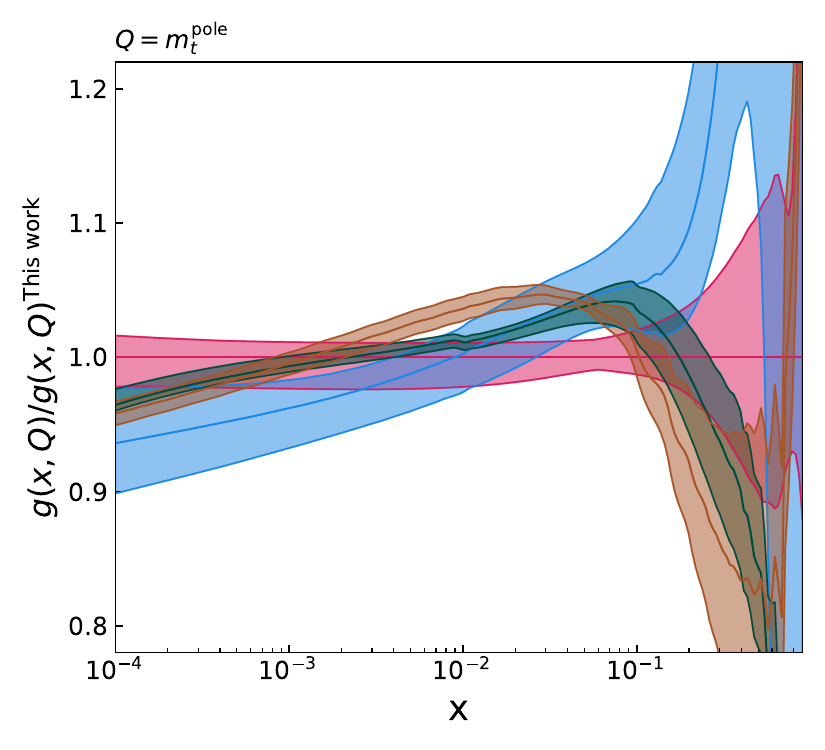}
    \includegraphics[width=0.44\linewidth]{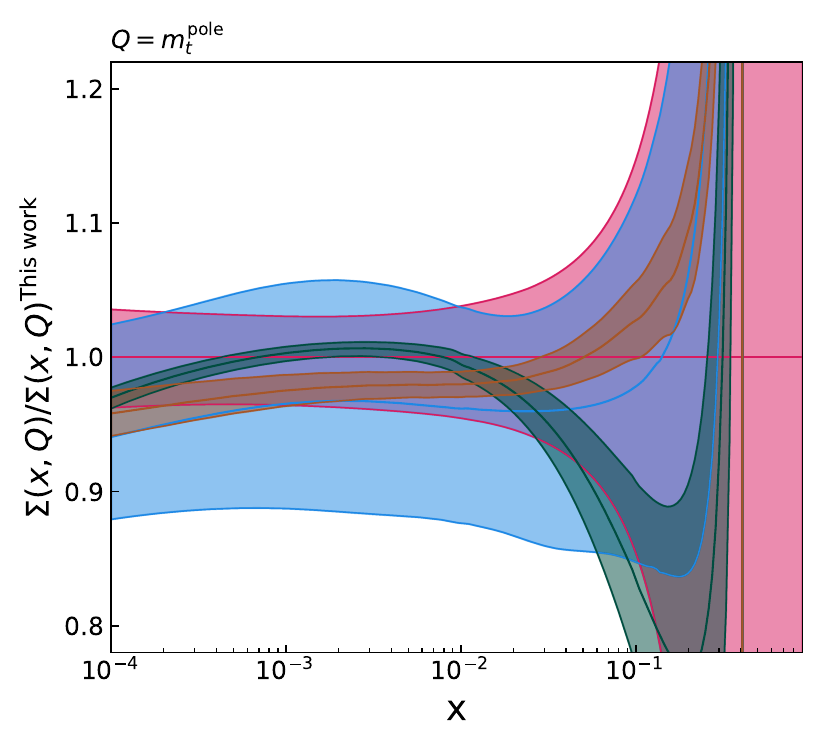}
    \caption{The $u$-valence (upper left), $d$-valence (upper right), gluon (lower left), and singlet (lower right) PDFs as a function of $x$ shown at the scale of $m_t^{\rm pole}$. The PDFs obtained in our analysis are presented with their total uncertainties and compared to the results of CT18 NNLO~\cite{Hou:2019efy}, HERAPDF2.0~\cite{H1:2015ubc} and ABMP16 NNLO~\cite{Alekhin:2017kpj}, shown in different colours, using \texttt{LHAPDF}~\cite{Buckley:2014ana}.}
\label{fig:pdfs_ref_qmt}
\end{figure}

The extracted PDFs are shown in Figure~\ref{fig:pdfs_ref_qmt}, at the scale of $m_t^{\rm pole}$, and are compared to the reults of CT18~\cite{Hou:2019efy}, HERAPDF2.0~\cite{H1:2015ubc} and ABMP16~\cite{Alekhin:2017kpj} PDF fits. Our results are in good agreement with the global PDFs for the quark distributions across the entire \(x\) range. For the gluon distribution, while the general shape is similar, our result shows an enhancement at low $x$. This enhancement is attributed to the larger starting scale used in this analysis reported in previous studies~\cite{xFitterDevelopersTeam:2018hym}.
The valence obtained in the HERAPDF2.0 fit, however,  shows quite different behaviour with respect to our result and the global PDFs.

The values of the SM parameters, obtained in our fit with the breakdown of uncertainties are summatized in Table~\ref{tab:sm_breakdown}.

\begin{table}[htbp]
\centering
\setlength{\tabcolsep}{5pt}
\begin{tabular}{|l|ccccc|c|}
\hline
\text{Parameter} & \text{Central} & \text{Exp.} & \text{Model} & \text{QCD scale} & \text{NRQCD scale} & \textbf{Total} \\
\hline
$m_t^{\text{pole}}$ (GeV) & 172.59 & $^{+0.35}_{-0.34}$ & $^{+0.04}_{-0.04}$ & $^{+0.35}_{-0.34}$ & $^{+0.12}_{-0.08}$ & $172.59^{+0.51}_{-0.49}$ \\
$\alpha_s(m_Z)$ & 0.1179 & $^{+0.0007}_{-0.0007}$ & $^{+0.00025}_{-0.00065}$ & $^{+0.0029}_{-0.0013}$ & $^{+0.0001}_{-0.0001}$ & $0.1179^{+0.0030}_{-0.0016}$ \\
$\sin^2\theta_{\text{eff}}^{\ell}$ & 0.23142 & $^{+0.00023}_{-0.00023}$ & $^{+0.00022}_{-0.00004}$ & $^{+0.00008}_{-0.00018}$ & $<0.00001$ & $0.23142^{+0.00032}_{-0.00029}$ \\
\hline
\end{tabular}
\caption{Breakdown of uncertainties for the extracted SM parameters. The total uncertainty is the quadrature sum of the individual components. }
\label{tab:sm_breakdown}
\end{table}
In our extraction, the MHOU is the dominant source of uncertainty for all three parameters. Large differences in the results caused by variation of the scales are driven primarily by the low-\(\sqrt{s}\) inclusive jet data, which are particularly sensitive to the light-quark distributions at NNLO. Future inclusion of measurements more sensitive to the sea and the $d$- valence could help to reduce this uncertainty. It is worth noting that the envelope approach for the QCD scale uncertainties is rather conservative. For future analyses, these scale variations could be incorporated as theory nuisance parameters
%~\cite{Tackmann:2024kci} 
or via a theory covariance matrix added to the experimental one~\cite{NNPDF:2024dpb}. For \(\alpha_s(m_Z)\) and \(\sin^2\theta_{\text{eff}}\), the model uncertainty is comparable to the experimental uncertainty, particularly the down variation of the strangeness fraction, \(f_s\).

\begin{figure}[htbp]
    \centering
    \includegraphics[width=0.32\linewidth]{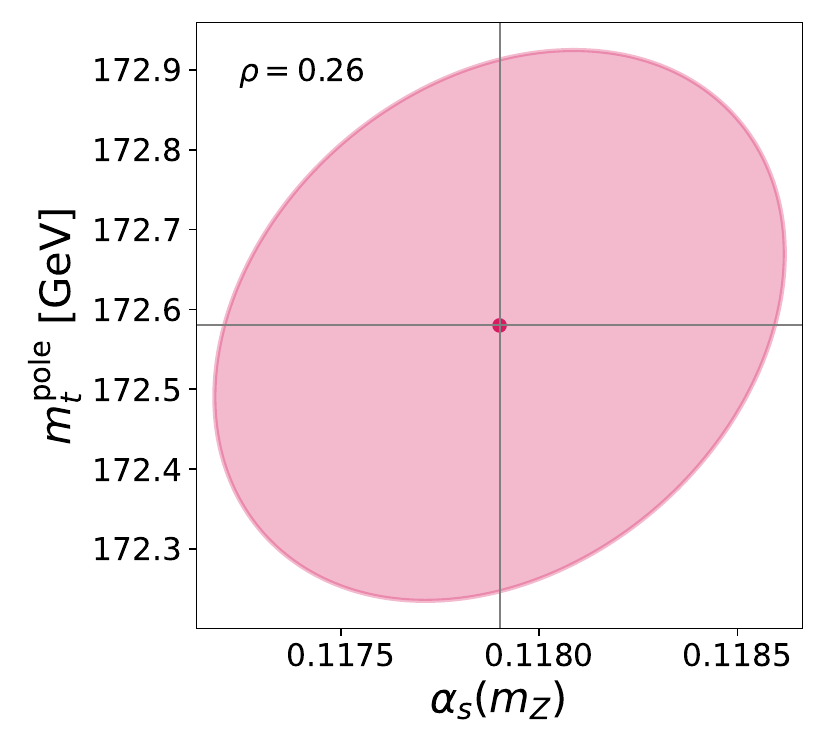}
    \includegraphics[width=0.32\linewidth]{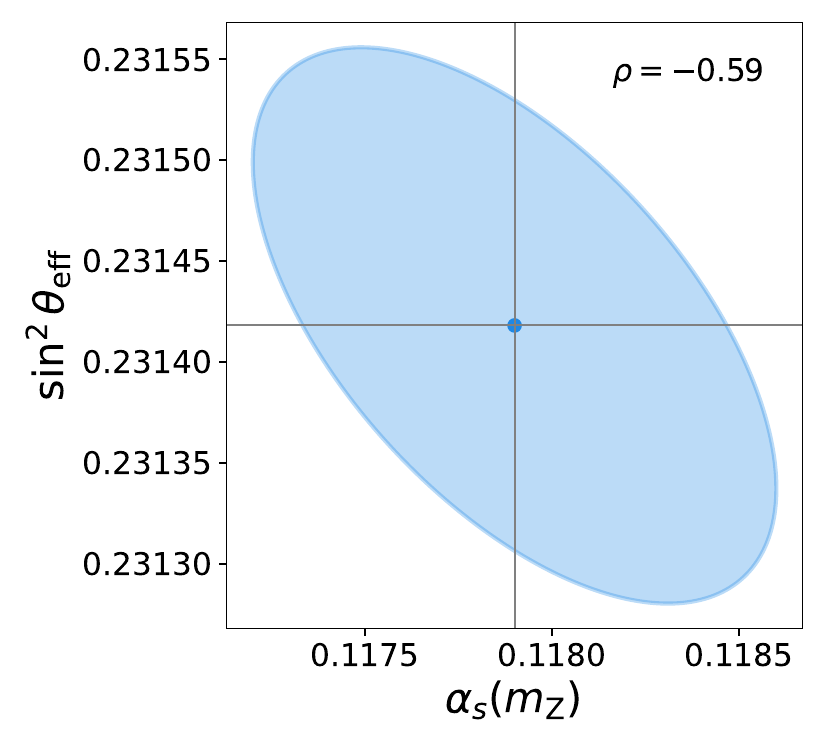}
    \includegraphics[width=0.32\linewidth]{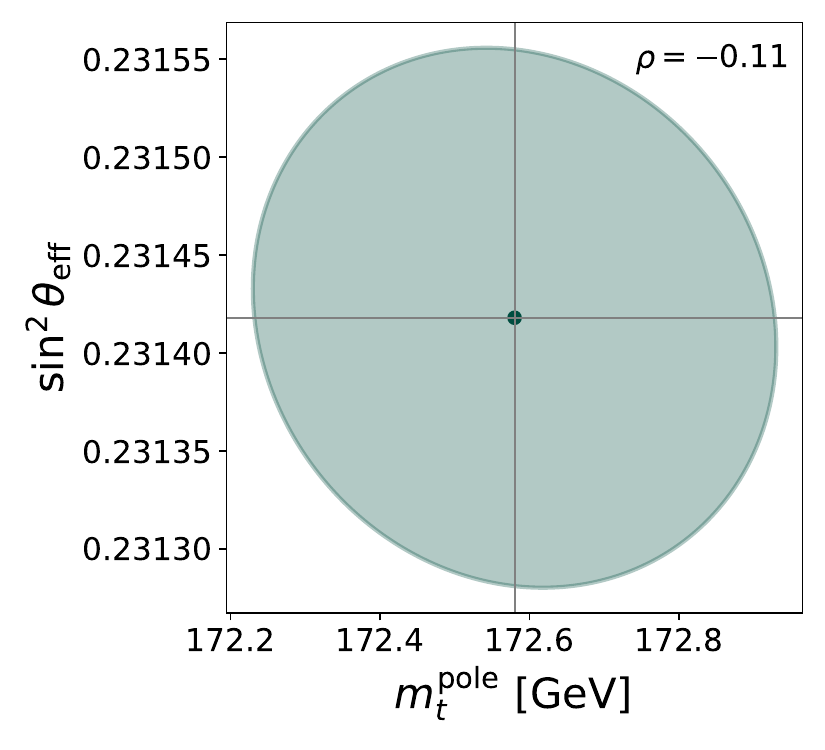}
    \caption{One standard-deviation correlation ellipses between the extracted SM parameters, with correlation coefficients of $\rho=0.26$ for \(m_t\) and \(\alpha_s(m_Z)\) (left), $\rho=-0.59$ for \(\alpha_s(m_Z)\) and \(\sin^2\theta_{\text{eff}}\) (middle), and $\rho=-0.11$ for \(m_t\) and \(\sin^2\theta_{\text{eff}}\) (right).}
    \label{fig:correlations}
\end{figure}

 The correlation between the extracted parameters is shown in Figure~\ref{fig:correlations}. We find a mild positive correlation between \(m_t\) and \(\alpha_s(m_Z)\), \(\rho = 0.26\). A moderate anti-correlation is observed between \(\alpha_s(m_Z)\) and \(\sin^2\theta_{\text{eff}}\), \(\rho = -0.59\), which explains why the QCD scale uncertainties and the down variation of \(f_s\) affect these two parameters in opposite directions. The correlation between \(m_t\) and \(\sin^2\theta_{\text{eff}}\) is found to be negligible, \(\rho = -0.11\).

To assess the impact of NRQCD and soft-gluon resummation, we compare the results of the FO extraction, where both effects are omitted, with the FO+NRQCD and FO+NRQCD+NLL results. 
The results of the three fits are summarised in Table~\ref{tab:sm_results}.

\begin{table}[htbp]
    \centering
    \setlength{\tabcolsep}{4pt} % Default is 6pt
    \begin{tabular}{|l|c|c|c|}
    \hline
         & \(m_t^{\text{pole}}\) [GeV] & \(\alpha_s(m_Z)\) & \(\sin^2\theta_{\text{eff}}^{\ell}\) \\ \hline
       FO NNLO  & 172.03  & 0.1178 & 0.23145   \\ 
       FO NNLO + NLO NRQCD & 172.48 & 0.1179 & 0.23143 \\
       FO NNLO + NLO NRQCD + NLL + EFT & 172.80 & 0.1175 & 0.23153  \\ \hline
       FO NNLO + NLO NRQCD + NLL & \(172.59^{+0.51}_{-0.49}\) & \(0.1179^{+0.0030}_{-0.0016}\) & \(0.23142^{+0.00032}_{-0.00029}\)  \\ \hline
    \end{tabular}
\caption{Results of the SM parameter extractions at FO NNLO and considering NLO NRQCD and resummation. The last row repeats the result of the full analysis with its total uncertainty for convenient comparison. Here, FO+NRQCD+NLL+EFT denotes the addition of NLO NRQCD, NLL soft-gluon threshold resummation effects, and EFT operators.}
\label{tab:sm_results}
\end{table}

The effect of NRQCD on \(m_t\) is significant, resulting in a shift of \(\Delta m_t = +0.45\) GeV. The effect on \(\alpha_s(m_Z)\) is \(\Delta \alpha_s(m_Z) = -0.00011\), is comparable to the experimental uncertainty. The additional inclusion of NLL soft-gluon resummation further shifts \(m_t\) by \(+0.11\) GeV, while its effect on \(\alpha_s(m_Z)\) and \(\sin^2\theta_{\text{eff}}\) is negligible. 

In the QCD+SMEFT version of the fit,  the Wilson coefficients    
$c_{tG}/\Lambda^2=0.08\pm0.08$ TeV$^{-2}$ and $c_{tq}^{(8)}/\Lambda^2=-0.4\pm 1.4$ TeV$^{-2}$ are obtained, consistent with the results of Ref.~\cite{Shen:2024sci} and the SM. Additional flexibility in the QCD+SMEFT fit is reflected in the changes of PDF and the SM parameters, although yet within experimental uncertainties, as reported in Table~\ref{tab:sm_results}.  

\section{Conclusions}
\label{sec:conclusions}

In this paper, the simultaneous extraction of PDFs together with \(\sin^2\theta_{\text{eff}}\), \(\alpha_s(m_Z)\), and \(m_t^{\text{pole}}\) is presented. The analysis is performed at NNLO in QCD using the \texttt{xFitter} framework, incorporating a HERA DIS data together with a set of precision LHC measurements providing particular sensitivity to the parameters of interest, and accounting for non-perturbative and threshold effects through a dedicated NRQCD calculation. The values obtained are competitive with state-of-the-art individual extractions, demonstrating the feasibility and importance of a fully consistent global analysis, considering NRQCD and threshold resummation effects. The observed correlations between the SM parameters underscore the need for a simultaneous approach to fully exploit the precision of future LHC data. This work paves the way for future analyses including higher-order corrections and a broader set of SMEFT operators.

\newpage
\acknowledgments
This work is supported by the Initiative and Networking Fund of the Helmholtz Association under grant number W2/3-123. The work by G.~L. is supported by A.v.Humboldt foundation. The work of O.~Z. has been supported by the \emph{MSCA4Ukraine Programme} of the European Commission through the Alexander von Humboldt foundation.

% Bibliography

%% [A] Recommended: using JHEP.bst file
 \bibliographystyle{JHEP}
 \bibliography{biblio.bib}

%% or
%% [B] Manual formatting (see below)
%% (i) We suggest to always provide author, title and journal data or doi:
%% in short all the informations that clearly identify a document.
%% (ii) please avoid comments such as "For a review'', "For some examples",
%% "and references therein" or move them in the text. In general, please leave only references in the bibliography and move all
%% accessory text in footnotes.
%% (iii) Also, please have only one work for each \bibitem.

%\begin{thebibliography}{99}

%\bibitem{a}
%Author,
%\emph{Title},
%\emph{J. Abbrev.} {\bf vol} (year) pg.

%\bibitem{b}
%Author,
%\emph{Title},
%arxiv:1234.5678.

%\bibitem{c}
%Author,
%\emph{Title},
%Publisher (year).
%\end{thebibliography}
\end{document}